\documentclass[useAMS]{mn2e}
\usepackage{lscape}
\usepackage{rotating}
\usepackage{amssymb}
\usepackage{float}

\def\msun{\hbox{M$_\odot$}}

\def\frac{\hbox{f$_{\rm mix}$}}

\def\t4{\hbox{t$_{\rm 4}$}}

\def\cm3{\hbox{cm$^{-3}$}}
\input psfig.sty
\input epsf.sty
\usepackage{graphicx}
\usepackage{epsfig}
\title[Rapid stellar rotation in YMCs]
{A high fraction of Be stars in young massive clusters: evidence for a large population of near-critically rotating stars}
\author[Bastian et al.] {N. Bastian$^{1}$, I. Cabrera-Ziri$^{1,2}$, F. Niederhofer$^{3}$, S. de Mink$^{4}$, C. Georgy$^{5}$, \newauthor D. Baade$^{2}$, M. Correnti$^{3}$, C. Usher$^{1}$, M. Romaniello$^{2}$\\
$^{1}$Astrophysics Research Institute, Liverpool John Moores University, 146 Brownlow Hill, Liverpool L3 5RF, UK\\
$^{2}$European Southern Observatory, Karl-Schwarzschild-Stra\ss e 2, D-85748 Garching bei M\"unchen, Germany\\
$^{3}$Space Telescope Science Institute, 3700 San Martin Drive, Baltimore, MD 21218, USA\\
$^{4}$Astronomical Institute Anton Pannekoek, University of Amsterdam, PO Box 94249, NL-1090GE Amsterdam, the Netherlands\\
$^{5}$Geneva Observatory, University of Geneva, Maillettes 51, 1290, Sauverny, Switzerland\\
}
\date{Accepted. Received; in original form}
\pagerange{\pageref{firstpage}--\pageref{lastpage}}
\pubyear{2015}
\begin{document}
\maketitle
\label{firstpage}
\begin{abstract}
Recent photometric analysis of the colour-magnitude diagrams (CMDs) of young massive clusters (YMCs) have found evidence for splitting in the main sequence and extended main sequence turn-offs, both of which have been suggested to be caused by stellar rotation.  Comparison of the observed main sequence splitting with models has led various authors to suggest a rather extreme stellar rotation distribution, with a minority ($10-30$\%) of stars with low rotational velocities and the remainder ($70-90$\%) of stars rotating near the critical rotation (i.e., near break-up).  We test this hypothesis by searching for Be stars within two YMCs in the LMC (NGC~1850 and NGC~1856), which are thought to be critically rotating stars with decretion disks that are (partially) ionised by their host stars.  In both clusters we detect large populations of Be stars at the main sequence turn-off ($\sim30-60$\% of stars), which supports previous suggestions of large populations of rapidly rotating stars within massive clusters. 
\end{abstract}
\begin{keywords} galaxies - star clusters
\end{keywords}

\section{Introduction}
\label{sec:intro}

A number of recent studies have found split main-sequences (split-ms) within young massive clusters in the Large Magellanic Cloud (LMC - Milone et al. 2015, 2016, Bastian et al. 2016; Niederhofer et al. in prep., Correnti et al. in prep.).  Attempts to model the split-ms using different aged populations, CNO variation, or Fe-spreads have not resulted in consistent fits to the observations (e.g., Milone et al. 2015).  However, stellar rotation appears to provide satisfactory fits (e.g., D'Antona et al.~2015) which is in agreement with studies that have explained the observed extended main-sequence turnoffs (eMSTOs) within young and intermediate age clusters through stellar rotation effects (e.g., Bastian \& de Mink~2009; Brandt \& Huang~2015; Niederhofer et al.~2015b).

One potential complication with this explanation, however, is that in order to have discrete main-sequences within the young clusters, a rather extreme distribution of stellar rotation rates is required, namely a bi-modal distribution with a peak near zero velocity and another near the critical rotation\footnote{As in previous works, i.e. Granada et al.~(2013), we use critical rotation to mean the velocity required for the centrifugal force to counterbalance gravity at the equator.} ($\omega = \Omega_{\rm rot} / \Omega_{\rm critical} \sim 0.9$ - i.e., D'Antona et al.~2015).  In the massive clusters studied so far, the ``red ms" (interpreted as being due to the high $\omega$ stars) is the dominant one, containing $\sim67$\% or more of the stars within the magnitude interval where it is visible, although there is considerable scatter from cluster to cluster (e.g., Milone et al.~2015, 2016; Niederhofer et al. in prep.).  D'Antona et al.~(2015) have attempted to explain this bi-modal rotational distribution by assuming that all stars begin their lives rapidly rotating and stars in binary systems can be effectively braked due to tidal synchronization.  However, it is currently unexplained why stars within massive clusters would all begin their lives as rapid rotators (at least in the stellar mass range explored so far) as stars in the field or in the lower mass open clusters do not appear to contain large fractions of rapid rotators ($<10\%$ rapid rotators, e.g., McSwain \& Gies~2005).

In order to confirm or refute the existence of a large population of rapidly rotating stars within young massive clusters, the ideal way would be to spectroscopically measure rotation rates for large samples of stars, sampling both the 'blue ms' and 'red ms'.  An alternative method is to search for Be stars within the cluster, based on the presence of H$\alpha$ emission (e.g., Keller et al.~2000).  Be stars are rapidly rotating stars, with $\omega \ge 0.88-0.95$ (e.g., Townsend et al.~2004; Fr\'emat et al. 2005; Delaa et al.~2011; Meilland et al. 2012; Rivinius et al.~2013), that develop decretion discs that become ionised by the host star, hence appear in H$\alpha$ emission.  While this measurement is certainly more crude than the measurements of stellar rotation rates directly, it should be able to discern whether a large population of rapidly rotating stars exist within the clusters.

In the present work, we search for Be stars in two young massive clusters in the LMC, NGC~1850 and 1856, both of which have been observed to have split main sequences (Bastian et al.~2016; Niederhofer et al.~in prep; Correnti et al.~in prep. and Milone et al.~2015, respectively).  We use existing HST based photometric catalogues and archival data to search for stars with H$\alpha$ emission.   In \S~\ref{sec:obs} we present the catalogues used and in \S~\ref{sec:analysis} we present an analysis techniques.  Finally, in \S~\ref{sec:discussion} we discuss our results and their implications and present our conclusions. %while in \S~\ref{sec:summary} we summarise our results and present our conclusions.

\begin{table*} 

  \begin{tabular}
    {lcccccc} ID& log (Age/yr) & log (Mass/\msun) & Be star fraction & M$_{\rm TO}$ (\msun)$^{a}$. \\
%    & & & &  & mag & & \msun & \msun & \\ 
    \hline 
 NGC~1850 & 7.9 & 4.86 & 0.19-0.62$^{b}$ & $4.6-5$\\
NGC~1856 & 8.45 & 4.88 & 0.33 & $2.9-3.1$\\
%;NGC~1866 & 8.12 (8.25) & 4.91 & 10.4 & 2.8 & 9.2\\
    \hline 
  \end{tabular}
\caption{Properties of the two LMC YMCs discussed in the present work.  Values taken from Niederhofer et al.~(2015a) and Bastian et al.~(2016).  In both clusters we are sensitive to H$\alpha$ down to $\sim2$~\msun.  $^{a}$The main sequence turn-off mass based on the SYCLIST models with Z=0.006 at the appropriate age for each cluster.  The range given shows the role of stellar rotation (with the low mass representing slow rotators and the higher end of the range representing $\omega=0.9$. $^{b}$The range given represents the Be star fraction from $19 > F438W > 16.5$.}
\label{tab:objects}
\end{table*}

\section{Observations, Catalogues and Methods}
\label{sec:obs}

\subsection{Catalogues}
For the present study we make use of previously published catalogues and archival data.  For NGC~1850, we use HST/WFC3 F336W and F438W photometry presented in Bastian et al.~(2016).  Additionally, we use the HST/WFPC2 images in F656N and F675W which were part of the GO-6101 programme (PI Gilmozzi) in order to search for stars with excess H$\alpha$ emission.  The WFPC2 images were reduced and the photometry carried out in the same way as described in Panagia et al.~(2000) and Romaniello et al.~(2002).  We focus on the inner $5.3$~pc (22") region of the cluster in order to avoid any contaminating H$\alpha$ emission from the ionised nearby regions.

For NGC~1856 we use the HST/WFC3 photometry presented in Correnti et al.~(2015), namely the F438W, F555W, F814W, and the F656N measurements.  As in Correnti et al.~(2015) we find that there exists differential extinction across the field of view.  We have corrected for this using the technique described in Milone et al.~(2012), however we note that correcting for differential extinction or not does not change our main conclusions.

\subsection{Methods}

\begin{figure*}
\centering
\includegraphics[width=8cm]{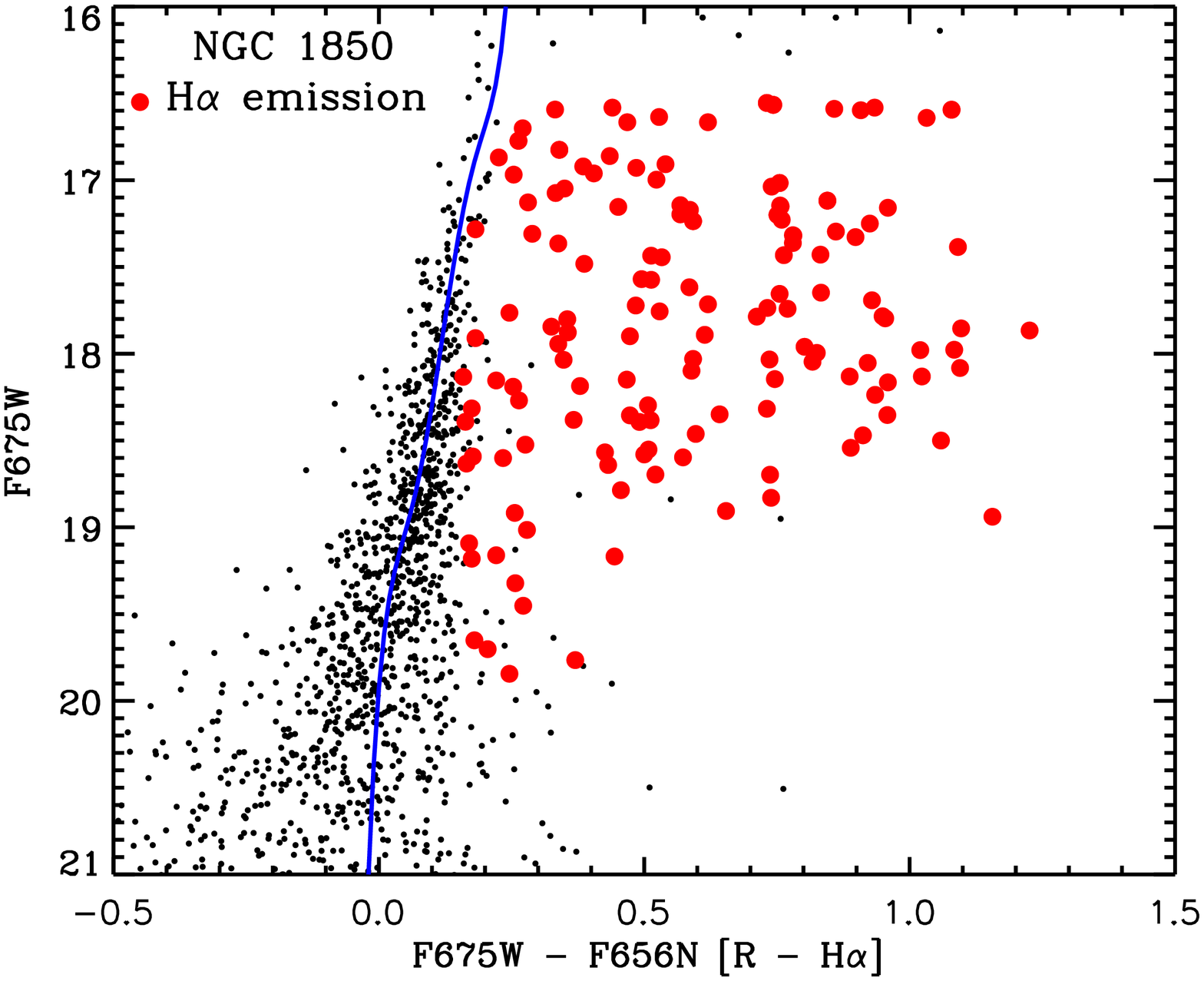}
\includegraphics[width=8cm]{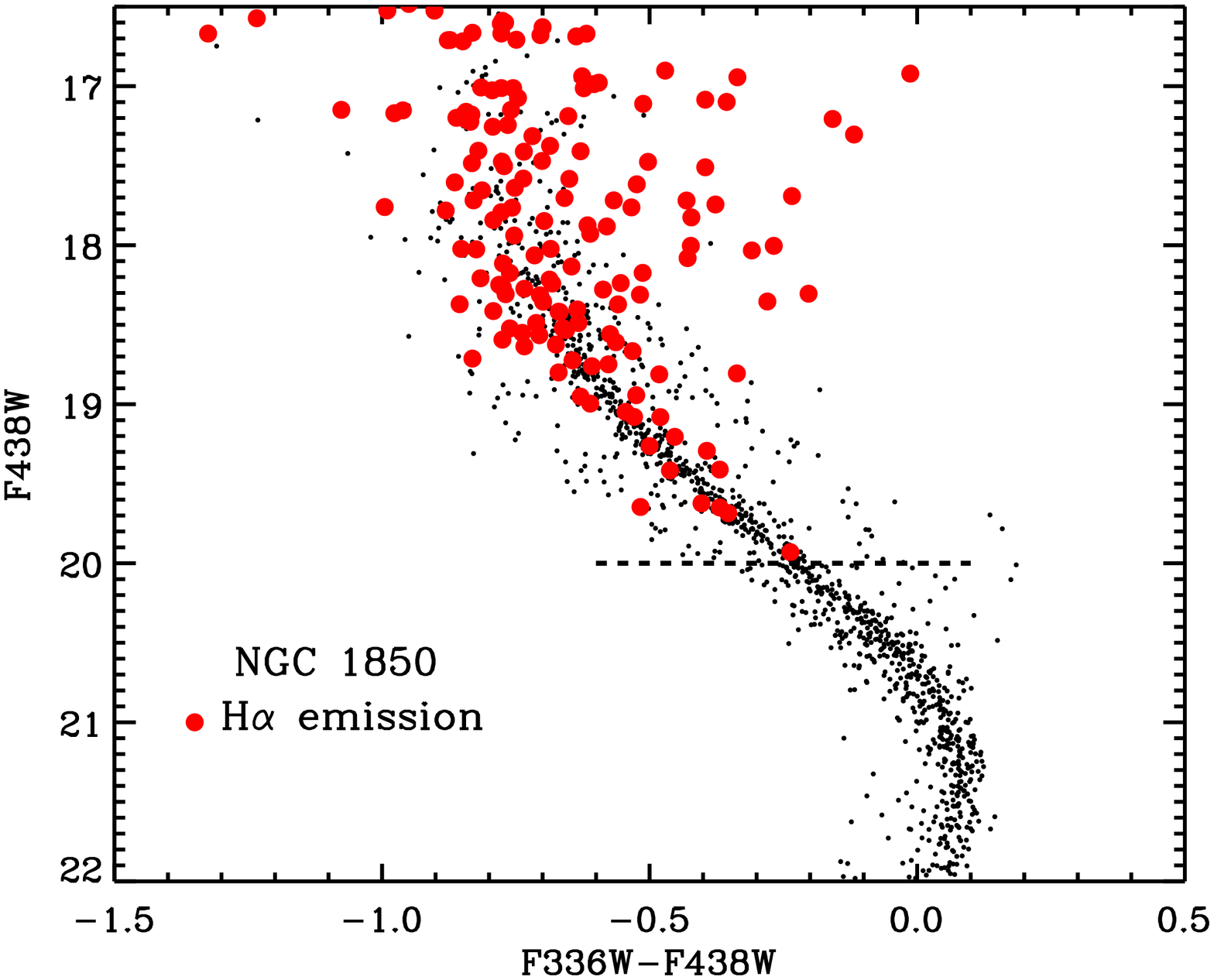}

\caption{{\bf Left:} A CMD of NGC 1850 used to identify stars with H$\alpha$ excess (marked as filled red circles).  The blue solid line shows the adopted ridge line of the stars not showing H$\alpha$ excess.  {\bf Right:} A broad band CMD of the same cluster with the stars with H$\alpha$ excess marked.  The horizontal dashed line denotes the estimate of the magnitude where we expect to be complete in the detection of H$\alpha$ excess.}
\label{fig:cmd_n1850}
\end{figure*}

\begin{figure*}
\centering
\includegraphics[width=8cm]{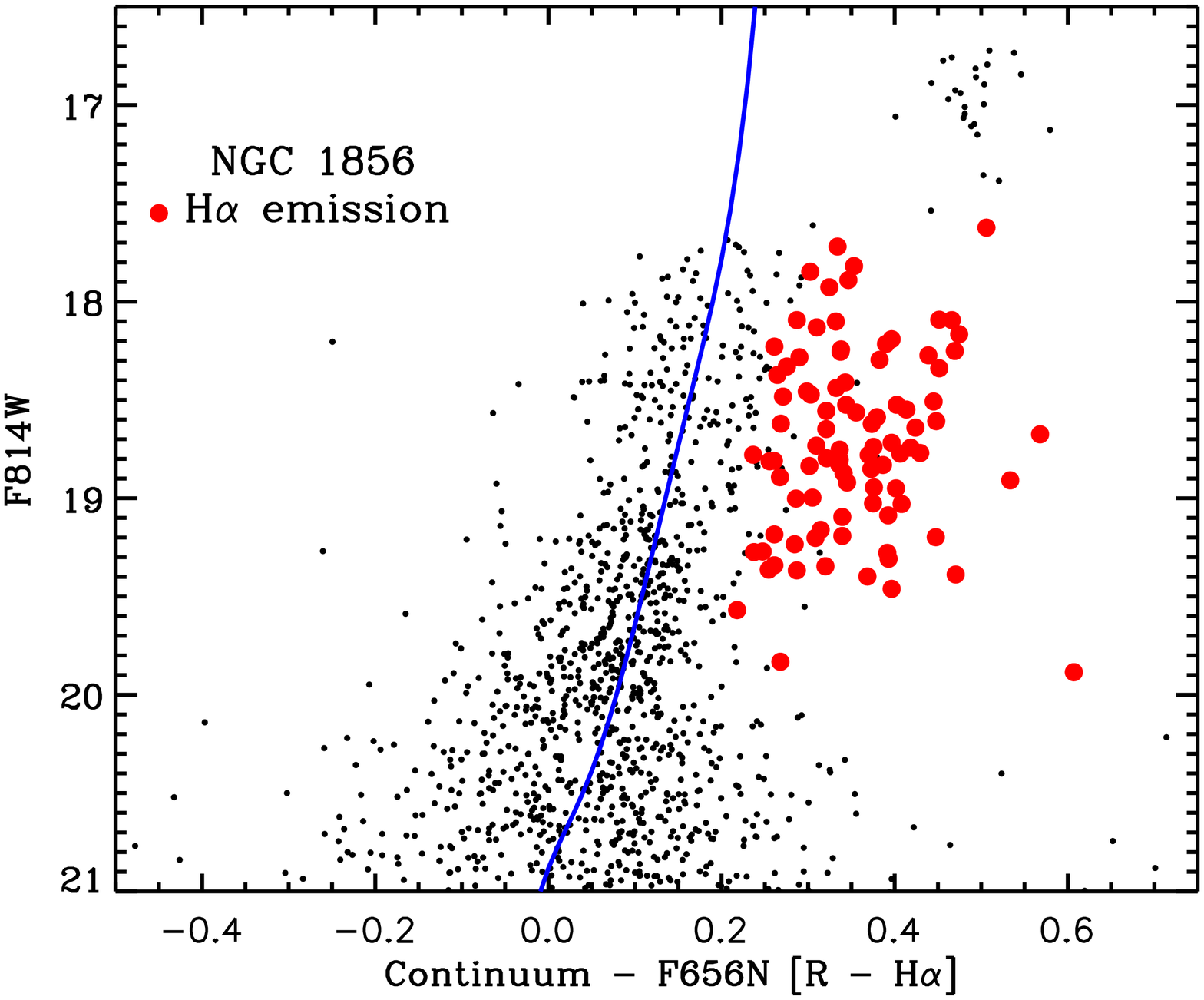}
\includegraphics[width=8cm]{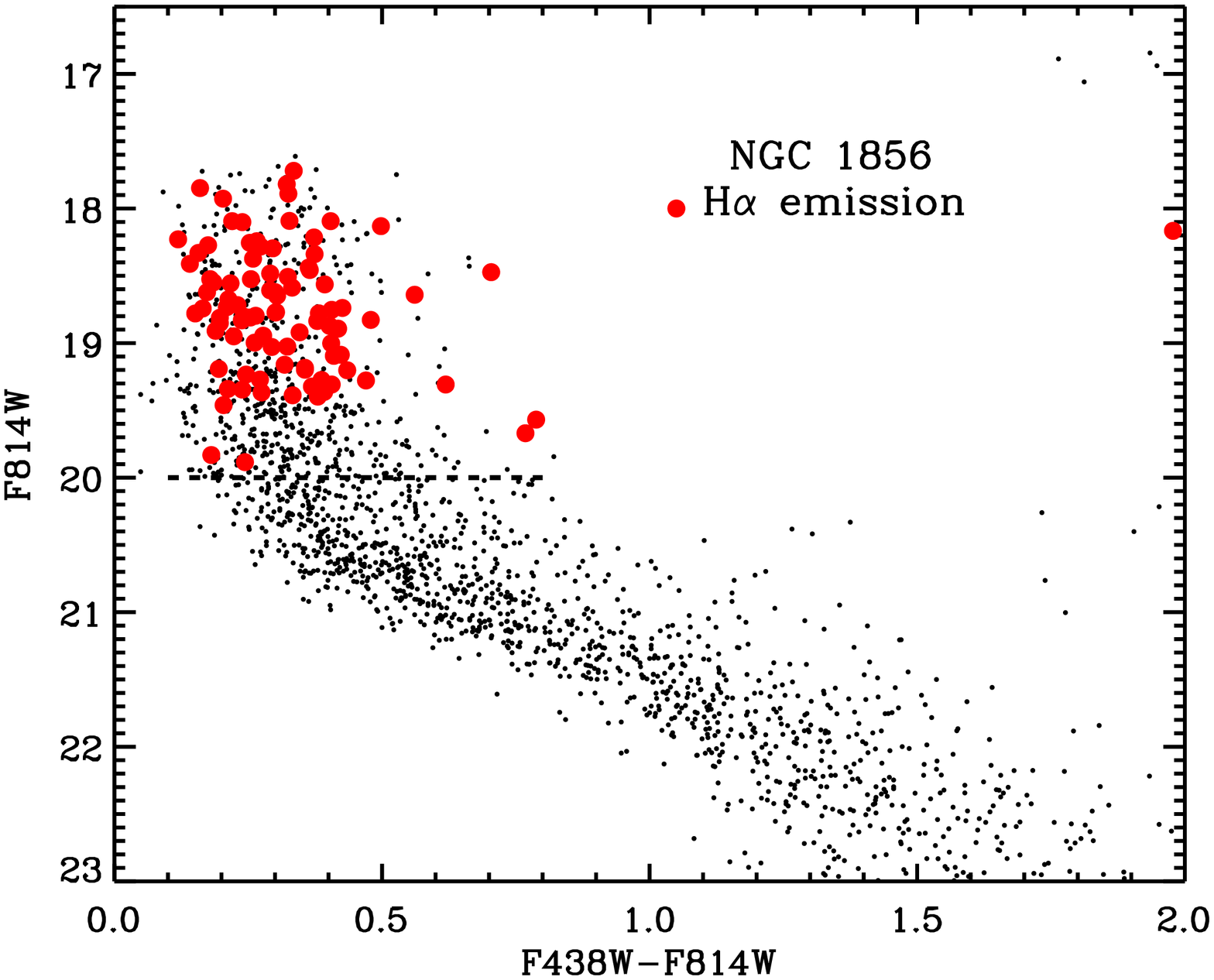}

\caption{The same as Fig.~\ref{fig:cmd_n1850}, but now for NGC~1856.}
\label{fig:cmd_n1856}
\end{figure*}

In order to search for stars with H$\alpha$ excess we follow the method of Keller et al.~(2000).  First, we make a colour-magnitude diagram (CMD) with the continuum flux of a star (e.g., either the F675W magnitude or a linear combination of nearby filters such as the F555W and F814W - see Zeidler et al.~2015) minus the F656N magnitude (which includes both the continuum and any H$\alpha$ emission that is present) vs. F675W (or a nearby filter).  Sources with H$\alpha$ excess scatter to the red in such a diagram.  Next we fit a polynomial to the locus of points without H$\alpha$ excess (i.e. the nominal ridge line).  Finally, we search for stars that are $3\sigma$ or more to the red from the ridge line, based on their estimated errors.  The selection of stars with excess H$\alpha$ emission is shown in the left panels of Figs.~\ref{fig:cmd_n1850} \& ~\ref{fig:cmd_n1856}, for NGC~1850 and NGC~1856, respectively.

In the right panels of Figs.~\ref{fig:cmd_n1850} \& ~\ref{fig:cmd_n1856} we show the position of the H$\alpha$ excess stars in the broad-band CMDs of the clusters.  The dashed horizontal lines indicate the magnitude at which we expect to be complete until, based on histograms of the H$\alpha$ magnitudes within the clusters.

Correnti et al.~(2015) used a similar approach (see also McSwain \& Gies~2005) for NGC~1856 to identify H$\alpha$ emitting stars, but they focussed their attention on fainter magnitudes (and redder broad-band colours) in order to search for pre-main sequence stars.  However, they also found a large population of H$\alpha$ emitting stars near the main sequence turn-off.

Both clusters host a large population of Be stars within them.  We note, however, that the method used here results in a lower limit of Be stars within the cluster, as the Be phase itself is transitory (e.g., Rivinius et al. 2013, and references therein) as Be stars can temporarily lose their line emission, i.e., their disk, completely (e.g., Baade et al.~1988).  However, the relative fraction of time stars spend in the emission or non-emission state is currently unknown, so we will use the observed values as lower limits.  

We note that the H$\alpha$ excess observed in stars in NGC~1850 is significantly larger than for NGC~1856.  This is likely due to the different ages of the clusters (NGC~1850 is $\sim80$~Myr while NGC~1856 is $\sim280$~Myr - e.g., Bastian \& Silva-Villa~2013), resulting in significant temperature differences between stars on the respective turnoffs.  For NGC~1850 we expect the turn-off stars to be $\sim5$~\msun, while for NGC~1856 the turnoff stars have masses of $\sim3$~\msun.  The difference in temperature, and subsequently ionising flux between the stars on the turnoff, mean that the H$\alpha$ excess is expected to be larger in NGC~1850, as observed.  In Table~\ref{tab:objects} we give the basic properties of the clusters and their stellar populations.

As a point of reference, the splitting of the main sequence in NGC~1850 can be seen just above and below the dashed horizontal line the right panel of Fig.~\ref{fig:cmd_n1850}.  This will be studied in more detail in Niederhofer et al. (in prep) and Correnti et al. (in prep.).  The split in the main sequence in NGC~1856 is not clearly visible in the right panel of Fig.~\ref{fig:cmd_n1856} due to the choice of filters, as it becomes more evident in ultraviolet filters (e.g., Milone et al. 2015, D'Antona et al.~2015).

%NGC~1850 - H$\alpha$ and MSTO CMD - Gilmozzi et al. (1994)
%For the blue MS - Niederhofer et al. (2016 in prep)

%NGC~1856 - H$\alpha$ and MSTO CMD - Correnti et al. (2015)

%NGC 1866 - blue MS - Niederhofer et al. (2016 in prep)

%Keller et al. (2000) - WFPC2 data for young clusters (15-25 Myr)

\section{Analysis}
\label{sec:analysis}

\subsection{NGC~1850}

As seen in Fig.~\ref{fig:cmd_n1850} a significant fraction of stars near the MSTO in NGC~1850 display H$\alpha$ emission.  We have quantified the fraction of stars with H$\alpha$ emission as a function of magnitude and find fractions of 0.04, 0.19, 0.40 and 0.62 for the magnitude ranges of $20 > F438W \ge 19$, $19 > F438W \ge 18$,  $18 > F438W \ge 17$,  $17 > F438W \ge 16.5$,  respectively.

A dependence of the fraction of H$\alpha$ emitting stars as a function of magnitude has been previously found (Keller et al. 2000; 2001). This may simply be due to the fact that brighter stars (at a fixed age) have higher temperatures, allowing for the discs to be more ionised and hence brighter in emission lines. Alternatively it may be that stars are more likely to possess discs as their approach the end of their MS lifetimes.  Another possible cause is that at the brightest luminosities, stars that are non-rapid rotators will have already left the MS, leaving only rapid rotators behind (which stay on the MS for longer due to rotational mixing).

We can compare the observed Be star fraction with that between the blue-MS and red-MS, which are the slow/non-rotating and rapidly rotating stars, respectively in the interpretation that their position is due to rotation (e.g., D'Antona et al.~2015).  The splitting in the MS can be seen in the right panel of Fig.~\ref{fig:cmd_n1850} just above the dashed horizontal line.  Niederhofer et al.~(in prep) show that the red-MS is the dominant population, making up $\sim92$\% at a given luminosity (where the two sequences can be clearly separated).  This is similar to that found in NGC~1755 (Milone et al.~2016), a similarly aged  ($\sim80$~Myr), although slightly lower mass ($\sim3\times10^4$~\msun) cluster, where the red-MS makes up $\sim75$\% of the population.  For NGC~1850, the high fraction of Be stars is consistent with the high fraction of red-MS stars, given that the Be star fraction is a lower limit.

Finally, we have created a synthetic cluster using the SYCLIST models (Georgy et al.~2014) made up of two populations at an age of 100~Myr, one that is made up of slow rotators ($\omega < 0.1$) and one where all stars are rotating at $\omega=0.9$, with the two populations having the same number of stars.  The results are shown in Fig.~\ref{fig:cmd_example}, where the two separate MSs can be seen. The inclination angles have been distributed randomly for both populations.  Note that the rapidly rotating stars are observed across the entire MSTO region and not just located on one side.  This is consistent with the observations (i.e., Fig.~\ref{fig:cmd_n1850}) and may make it difficult to directly associate Vsini measurements with the spread in the MSTO.  We note that these simulations do not include extinction to the stars from the decretion disc itself.  Hence, in the observations we would expect some of the rapidly rotating stars (those seen nearly edge on) to shift towards fainter magnitudes and redder colours, consistent with the observations (see Fig.~\ref{fig:cmd_n1850}).  The models also do not include include Balmer continuum emission (from the stars with strong Balmer line emission).
%at $3\sigma$\\
%$20 < b < 19$ - fraction = 0.039\\
%$19 < b < 18$ - fraction = 0.189\\
%$18 < b < 17$ - fraction = 0.404\\
%$17 < b < 16.5$ - fraction = 0.622\\

\subsection{NGC~1856}

Next we study the Be fraction in  NGC~1856.  As this cluster is significantly older than NGC~1850 ($\sim300$~Myr vs. $\sim100$~Myr) the stars on the MSTO have lower masses ($\sim3$~\msun), and hence have lower ionising fluxes.  Hence, some stars with discs may not be observable as Be stars, i.e. only stars with relatively strong emission will be detectable.

For stars brighter than $V=19.5$ we find a Be star fraction of 0.33.  Unlike NGC~1850, however, we do not find any significant change in the Be star fraction as a function of magnitude.  This may be due to the fact that at this mass the ionising flux from the host star changes less steeply with magnitude.

We can compare the observed Be star fraction with the ratio of blue (non-rotating) and red (rapidly rotating) stars on the MS.  In the magnitude range where the two sequences were readily distinguishable Milone et al.~(2015) found that the red MS dominates by number, making up $\sim2/3$ of the stars.  Note that in Fig.~\ref{fig:cmd_n1856} the splitting of the MS is not seen, which is due to the filter choice (the split is larger in the ultraviolet - e.g., Milone et al.~2015; D'Antona et al.~2015).  The range where the two sequences in the CMD can be differentiated is $\sim1.7 - 3$~\msun.  The observe Be star fraction of 0.33 is smaller than the fraction of the red MS ($\sim0.67$), however as discussed above, the Be star fraction is a lower limit to the number of rapid rotators in the cluster, and it is significantly higher than in typical open clusters or in the field (e.g., McSwain \& Gies 2005).  In NGC~1856, the fraction of stars on the red MS and the fraction of Be stars is lower than in NGC~1850.

The conclusion reached here is similar as that for NGC~1850, that a significant fraction of stars within the cluster are rotating near the critical rotation rate.

\begin{figure}
\centering
\includegraphics[width=8cm]{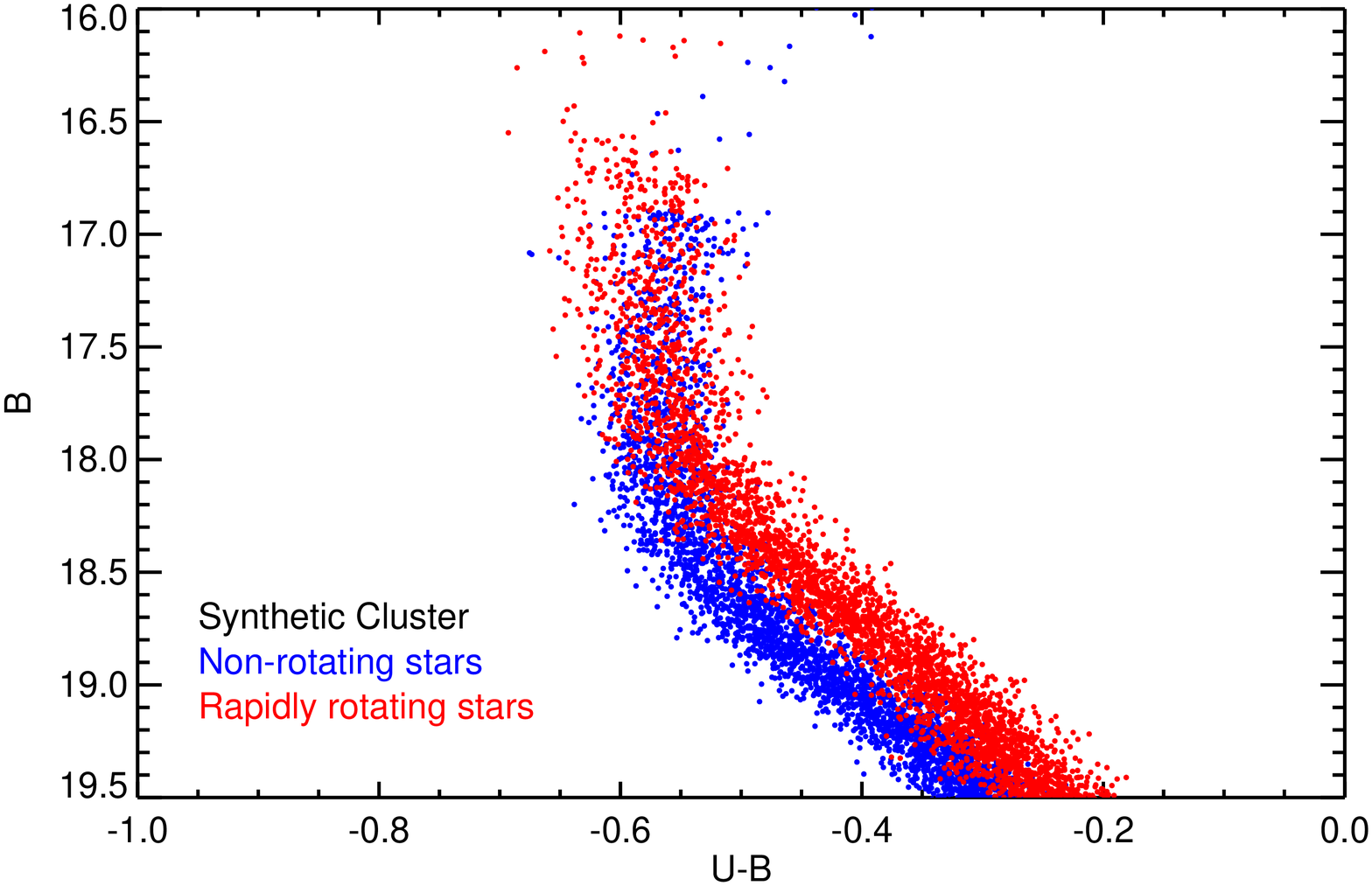}
\includegraphics[width=8cm]{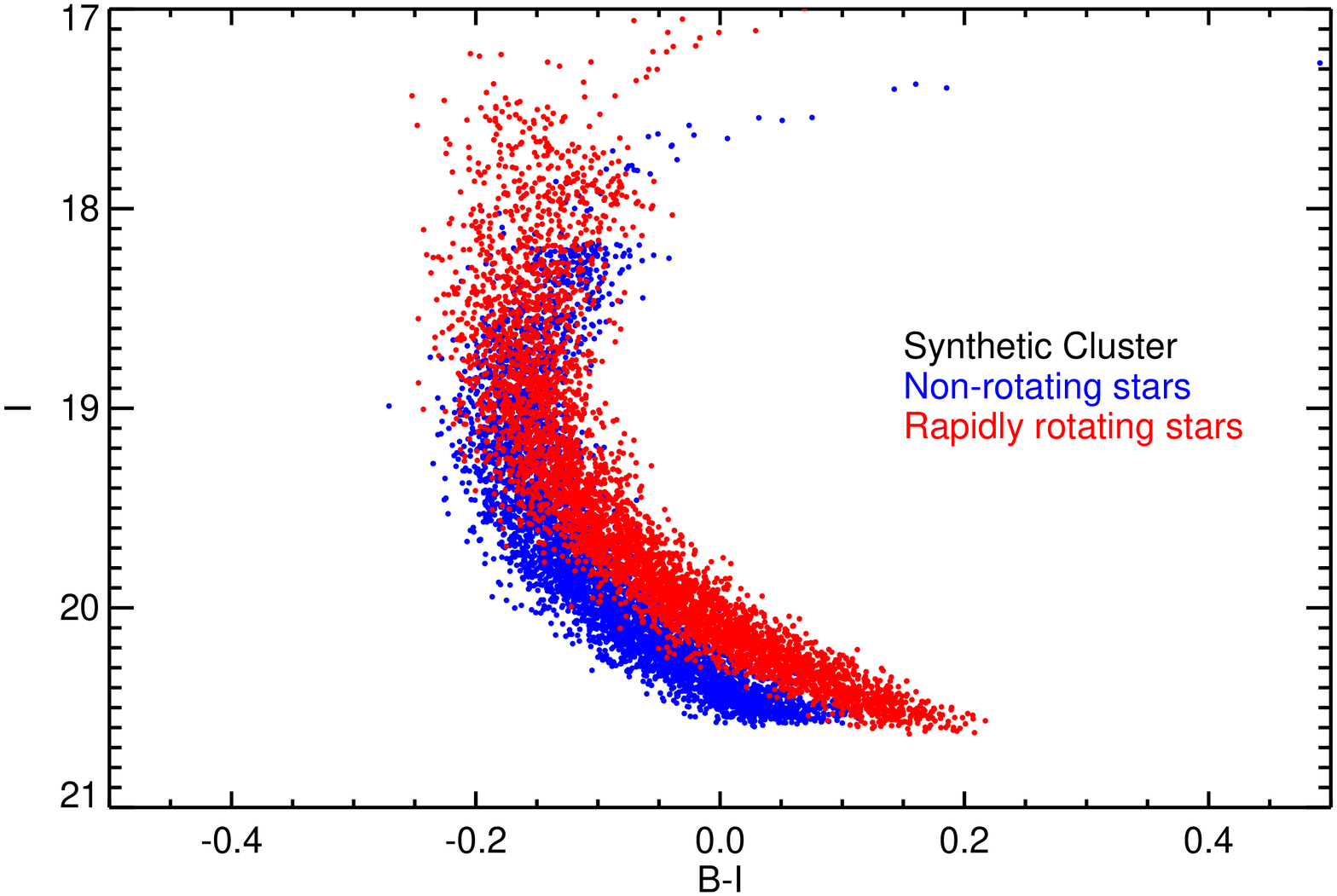}

\caption{{\bf Top:} A synthetic cluster made with the SYCLIST models with two populations that differ only in their rotation rates ($\omega=0$ and $\omega=0.9$) at an age of 100~Myr (meant to broadly reproduce NGC~1850).  Representative errors were taken from the observations of NGC~1850, and we have not included binaries.  Additionally, we have not taken into account reddening of stars that have decretion discs seen nearly edge on, nor have we included bound-free Balmer continuum emission from the disc itself.  Note that rapid rotators are expected to be present across the MSTO, and not just limited to one side.  {\bf Bottom:} The same but now for an age of 300~Myr, meant to broadly reproduce NGC~1856.  In both panels note that the filters are similar, but not identical to, the HST filters used in the present work. }
\label{fig:cmd_example}
\end{figure}

Fig.~\ref{fig:cmd_example} (bottom panel) shows a synthetic cluster made with the SYCLIST models for an age similar to NGC~1856 ($\sim300$~Myr).  As was found for NGC~1850 (top panel), the rapidly rotating stars are found across the MSTO, not just on a single side, consistent with the observed colour/magnitude distribution of the H$\alpha$ excess stars.

%NGC~1856 (Milone et al. 2015) where to red-MS makes up 2/3rd of the population

%Potential mass dependence, unlikely as there wasn't a strong age dependence in the previous works

\section{Discussion and Conclusions}
\label{sec:discussion}

%The exact rotation rate, relative to the critical rotation, that is required to form an decretion disc, not precisely known, due to systematic uncertainties in estimating the actual rotation rate in rapidly rotating stars.  However, the consensus is that a star must have $\omega > 0.75$ in order to become a Be star (REF).

Keller et al.~(2000) have used HST/WFPC2 photometry to search for Be stars within three LMC (NGC~1818, 2004, 2100) and one SMC (NGC~330) clusters with ages between $\sim15$ and $\sim25$~Myr.  They found Be star fractions of $10-15$\% for stars brighter than $V=19$, similar to, although slightly lower than found in NGC~1850 and NGC~1856 in the current study.  However, they also found that the Be star fraction increases substantially towards the MSTO, peaking at 0.35-0.5.  The authors note that these values are substantially higher than that found in the field, which is suggestive that stellar rotation (at least this evolutionary state) is dependent on the surrounding environment.  Martayan et al.~(2010) used slitless spectroscopy to search for Be stars in a sample of SMC clusters.  They also found that the Be star fraction depends on spectral type (luminosity), but that by spectral types of B5 or B6 (those at the MSTO in NGC~1850) the fraction of Be stars was below 5\%.  Hence, NGC~1850 and NGC~1856 have much higher Be star fractions than the field or lower mass clusters, when similar spectral types are compared.

As discussed previously, the dual MSs observed in a growing number of young massive clusters in the LMC/SMC is likely due to a bi-modal rotational distribution, with one peak at lower rotation rates and another near the critical rotation rate.  Such a large population of rapidly rotating stars would appear to be unique amongst stars within massive clusters.  The observations of large fractions of Be stars within such clusters lends support for this interpretation (i.e. of the presence of a large population of rapidly rotating stars within clusters).

It is worth noting that Dufton et al.~(2013) found a bi-modal rotational distribution of early-type B stars in the young region 30 Doradus in the LMC.  Hence, such distributions appear at all ages ($<10$~Myr in 30 Doradus to $\sim300$~Myr for NGC~1856) which implies that it is imprinted at birth (see also - see also Martayan et al.~2007) and is not an evolutionary feature (i.e. binary interactions slowing down rapid rotators to form the peak at low rotational periods - e.g., D'Antona et al.~2015).  

While a consensus on the minimum rotation rate required for Be stars to form has not been fully reached, the majority of estimates place this value at $\omega_{\rm Be} \approx 0.8$ or higher (c.f. Rivinius et al.~2013 and references therein).  This is in agreement with the rapid rotation rates required by D'Antona et al.~(2015) to explain the dual MSs observed in NGC~1856 ($\omega \approx 0.9$).

However, we note that while Be stars are good tracers of rapidly rotating stars, the Be phenomenon is intermittent, so there is likely a population of rapidly rotating stars that we have missed with our method.  Hence, the observed fraction of stars that display H$\alpha$ emission is a lower limit to the actual number of Be stars within the clusters.

Huang \& Gies (2008) and Huang et al.~(2010) have argued that while the incidence of rapid stellar rotators amongst B stars is higher in clusters than in the field, this is due to the relative youth of the clusters in previous surveys, and that the field stars were on average older, meaning that they have had a longer time to spin down.  The two clusters studied here have densities more than an order of magnitude higher than those used in the Huang et al. study, and are significantly older than the clusters used in that work.  The fact that we find high fractions of rapidly rotating stars in these two relative old clusters leads us to conclude that stellar density does play a major role in setting the rotation rates of stars, at least near globular cluster type masses ($\sim10^5$~\msun).

We conclude that high fraction of Be stars within these two YMCs, which implies a high fraction of rapidly rotating stars, argues in favour of the interpretation that the observed main sequence splits in YMCs is due to a bi-modal rotation distribution, with a majority of stars rotating near the critical velocity (e.g., D'Antona et al.~2015).  Such high fractions of rapid rotators are not seen in the field (c.f. Keller et al.~2000; Rivinius et al.~2013), suggesting that the cluster environment may impact the angular momentum distribution of the stars within them, which in turn may affect their evolution.  Such a high fraction of rapid rotators would manifest in extended main sequence turn-offs in clusters, as often observed in young and intermediate age clusters (e.g., Milone et al.~2009; Niederhofer et al.~2015b).

A high fraction of rapid rotators would be unique in stellar clusters and shows that the environment in which a star forms may influence its further evolution.  A number of anomalies (i.e. chemical and morphological CMD features) have been observed in the ancient globular clusters, i.e., multiple populations.  If GCs are indeed the ancient analogues to YMCs forming locally (see Kruijssen~2015 for a recent summary) we may expect them to share the rotation distribution properties observed in clusters like NGC~1850 and 1856.  Such potential stellar evolutionary means to form multiple populations (to explain why they are only found in GCs and not in the field in large numbers) may be an attractive avenue for future research as all scenarios put forward so far, mostly based on multiple generations of star-formation taking place within clusters, have significant problems and are all but ruled out (e.g., Bastian~2015).

In order to quantify the exact rotation rate distribution (Vsini) within clusters dedicated medium/high resolution spectroscopic surveys of cluster members are required.  The MSTO region of the CMDs of the two clusters presented here are well within reach of the capabilities of instruments like FLAMES/UVES on the VLT.

%\fbox{caveat:  if it is an evolutionary thing, the field will be lower}

%\subsection{A potential Link to Multiple Populations in globular clusters?}

%\section{Summary and Conclusions}
%\label{sec:summary}

%\vspace{-0.7cm}
\section*{Acknowledgments}

NB gratefully acknowledges financial support from the Royal Society (University Research Fellowship) and 
the European Research Council (ERC-CoG-646928, Multi-Pop).  F.N. gratefully acknowledges financial support for this project provided by NASA through grant HST-GO-14069 from the Space Telescope Science Institute, which is operated by the Association of Universities for Research in Astronomy, Inc., under NASA contract NAS526555. SdM acknowledges support by a Marie Sklodowska-Curie Action (H2020 MSCA-IF-2014, project id 661502).

\bsp
\label{lastpage}
\end{document}